\documentclass[showpacs,preprintnumbers,amsmath]{revtex4}
\usepackage{psfig}

\begin{document}

\title{Collision Rates in Charged Granular Gases}

\author{T.~Scheffler}
\author{D.E.~Wolf}
\email{d.wolf@uni-duisburg.de}

\address{Institut f\"ur Physik, Gerhard--Mercator--Universit\"at
  Duisburg, D-47048 Duisburg, Germany}

\date{\today}

\begin{abstract}
The dissipation rate due to inelastic collisions between equally
charged, insulating particles in a granular gas is calculated.  
It is equal to the known dissipation rate for uncharged
granular media multiplied by a Boltzmann-like factor, that originates
from Coulomb repulsion.  Particle correlations lead to
an effective potential that replaces the bare Coulomb
potential in the Boltzmann factor. Collisional cooling in a granular
gas proceeds with the known $t^{-2}$-law, until the kinetic energy of
the grains becomes smaller than the Coulomb barrier. Then the granular
temperature approaches a time dependence proportional to $1/\ln{t}$.
If the particles have different charges of equal sign, the collision 
rate can always be lowered by redistributing the charge, until all
particles carry the same charge. 
Finally granular flow through a vertical pipe is briefly discussed.
All results are confirmed by computer simulations. 
\end{abstract}

\pacs{PACS numbers: 45.70.-n, 05.20.Dd, 82.20.Pm}

\maketitle

\section{Introduction}

The particles in most granular materials carry a net electrical
charge. This charge emerges naturally due to contact electrification
during transport or is artificially induced in industrial processes.
It is well known~\cite{Kanazawa,Nieh88,Singh85}, for instance, that
particles always get charged when transported through a pipe. In industry,
contact electrification is used for dry separation of different
plastic materials or salts~\cite{Inculet}, which tend to get
oppositely charged and hence are deflected into opposite directions
when falling through a condensor. Another application is powder
varnishing, where uniformly charged pigment particles are blown
towards the object to be painted, which is oppositely charged.
Opposite charges are also used to cover one kind of particles with
smaller particles of another kind in order to reduce van-der-Waals cohesion 
e.g. for inhalable drugs.

Whereas the dynamics of electrically neutral grains has been studied
in great detail, little is known about what will change, if the grains
are charged. In this paper we present the answer for collisional
cooling, a basic phenomenon, which is responsible for
many of the remarkable properties of dilute granular media.
By collisional cooling one means that the relative motion of the grains,
which lets them collide and can be compared to the thermal
motion of molecules in a gas, becomes weaker with every
collision, because energy is irreversibly transferred to the internal
degrees of freedom of the grains.

In 1983 Haff~\cite{Haff} showed, that the rate, at which the kinetic
energy of the relative motion of the grains is dissipated in a homogeneous
granular gas, is proportional to $T^{3/2}$, where $T$ is the so called
granular temperature. It is defined as the mean square
fluctuation of the grain velocities divided by the space dimension:
\begin{equation}
T = \langle \vec v\,^2 - \langle \vec v \rangle^2 \rangle/3 .
\label{eq1}
\end{equation}
A consequence of this dissipation rate is that the granular
temperature of a freely cooling granular gas decays with time as
$t^{-2}$.  We shall discuss, how these laws change,
if the particles are uniformly charged. 

Due to the irreversible particle interactions large scale patterns
form in granular media, such as planetary rings \cite{Planets} or the
cellular patterns in vertically vibrated granular layers
\cite{Umbanhowar,Shinbrot,Eggers}.  
This happens even without external driving
\cite{Goldhirsch,McNamara,Luding99}, where one can distinguish a
kinetic, a shearing and a clustering regime. The regimes 
depend on the density, the system size and
on the restitution coefficient $e_{\rm n}=-v_{\rm n}'/v_{\rm n}$, which
is the ratio of the normal components of the relative velocities
before and after a collision between two spherical grains.
The $T^{3/2}$ cooling law holds, provided the restitution coefficient
may be regarded as independent of $v_{\rm n}$ \cite{Brilliantov}, and
the system remains approximately homogeneous \cite{Luding98}. The
latter condition defines 
the kinetic regime, which is observed for the highest values of the
restitution coefficient, whereas the two other regimes are more
complicated because of the inhomogeneities.  Such inhomogeneities can
only occur as transients, if all particles are equally charged,
because the Coulomb repulsion will homogenize the system again.

In order to avoid additional dissipation mechanisms due to eddy
currents within the grains we consider only insulating materials.
Unfortunately, up to now, no consistent microscopic theory for contact
electrification of insulators exists~\cite{Lowell}.  In powder
processing two types of charge distribution are
observed~\cite{Singh85}: A bipolar charging, where the charges of the
particles in the powder can have opposite sign and the whole powder is
almost neutral. The other case is monopolar charging, for which the
particles tend to carry charges of the same sign and the 
countercharge is transferred to the container walls.
It depends largely on the type
of processing, whether one observes bipolar or monopolar charging,
which means, that the material of the container, the material of the
powder and other more ambiguous things, like air humidity or room
temperature are important~\cite{Lowell}. 


The outline of this article is as follows: The next section specifies
the model we are considering. A simple derivation of the
dissipation rate in dilute charged granular media based on kinetic gas
theory is given in Sec.~\ref{sec_ANALY}.  Sec.~\ref{sec_DELTA}
compares the analytic results with computer simulations.  We find that
in non-dilute systems the Coulomb repulsion is effectively reduced.
This reduction will be explained, and we determine its dependence
on the solid fraction of the granular gas. Then we apply these results
to collisional cooling in Sec.~\ref{sec_COOL}. Sec.~\ref{sec_APPL}
addresses several aspects which are important for the applicability of 
the results for the dissipation rate: Are they valid locally in
inhomogeneous systems, e.g. in the presence of container walls?
What are the implications for granular flow through a vertical pipe?
How does the dissipation rate change, if the particles carry different
charges of equal sign? Finally, in Sec.~\ref{sec_DISC} we summarize
our results. In appendix A we explain the simulation method we
developed for this investigation \cite{Scheffler_MD}. 

\section{The model}

In this paper, we consider monopolar charging, which is the usual case
if insulators are transported through a metal
pipe~\cite{Nieh88,Singh85}. For simplicity we assume, that
all particles have the same point charge $q$ centred in a sphere of
diameter $d$ and mass $m$.  No polarization and no charge transfer
during contact will be considered. The particle velocities are assumed
to be much smaller than the velocity of light, so that relativistic
effects (retardation and magnetic fields due to the particle motion)
can be neglected.
The electrodynamic interaction between the particles can then be
approximated by the Coulomb potential:
\begin{equation}\label{Coulomb}
\Phi _{ij} = q^2 / r_{ij},
\end{equation}
where $r_{ij}$ is the distance between the centers of particles
$i$ and $j$. 

We consider the collisions as being instantaneous, which is a good
approximation for the dilute granular gas, where the time between
collisions is much longer than the duration of a contact between two
particles. As the incomplete restitution ($e_{\rm n}<1$) is the main 
dissipation
mechanism in granular gases, Coulomb friction will be neglected in this
paper. Also, the dependence of the restitution coefficient on the 
relative velocity \cite{Schaefer2,Brilliantov} will be ignored, so that
the constant $e_{\rm n}$ is the only material parameter in our model.

The particles are confined to a volume $V = L^3$ with periodic
boundary conditions in all three directions. The pressure in the
system has two positive contributions, one kinetic and one due to the
Coulomb repulsion. Although there are no confining walls, the system
cannot expand because the periodic boundary conditions mean
topologically, that the particles are restricted to a hypertorus of
fixed volume. As we are going to show in Sec.\ref{Walls},
the periodic volume can be thought of as a sufficiently homogeneous subpart
of a larger system, which is kept from expanding by reflecting walls.

For vanishing particle diameter this model corresponds to
the One Component Plasma (OCP)~\cite{Baus}. In the OCP a classical
plasma is modelled by positive point charges (the ions) acting via the
Coulomb potential, whereas the electrons are considered to be
homogeneously smeared out over the whole system. In the OCP the
electron background cannot be polarised, i.e. Debye screening does not exist,
as is the case in our model, too.  

In order to take the periodic images of the simulation cell into
account we used an Ewald summation. It is only possible for the forces
-- which is all one needs in a molecular dynamics
simulation.\footnote{The potential energy summed over all periodic
  images diverges.} We checked that for our simulation parameters only the
nearest periodic images of a particle contribute noticably to the
force, so that most of our data could be obtained with the simple
minimum image convention instead of the Ewald summation.

\section{Analytical results for dilute systems} \label{sec_ANALY}

In this section we derive an approximate expression for the
dissipation rate in a dilute system of granular spheres, 
which interact with a purely repulsive, rotationally invariant pair
potential. The reasoning is a simple adaptation of the kinetic theory
of a hard sphere gas with inelastic collisions \cite{Lun84,Jenkins}.
In particular we assume that the pair potential is sufficiently well
behaved that correlations of the particles are unimportant in the
dilute limit, that the particle density remains uniform and that the
velocity distribution remains approximately Gau{\ss}ian in spite of
the nonstationarity due to the collisional cooling process.

It is one of the surprising results of the computer simulations
presented in the next section that these assumptions give the correct
high temperature behaviour even for the Coulomb interaction, which is
known not to be ``well behaved'' in the above sense because of its
long range \cite{Koga70}.  One reason for the applicability is that
the hard core repulsion sets an upper bound on the Coulomb interaction
$\Phi_{ij} \leq q^2/d$. Hence, if the kinetic energy of the particles
is much larger than this value, the collision rate will be
approximately the same as in the uncharged case. This is a crucial
simplification compared to charged point particles.  Using the
analytic form of the dissipation rate in the dilute limit derived
here, we will discuss the dissipation in a non-dilute system, where
correlations are important, in the next section.

We start with calculating the collision frequency of a fixed particle
$i$ with any of the other particles $j$. If they were not charged, two
particles would collide provided the relative velocity $\vec u $
points into the direction of the distance vector $\vec r = \vec r_j -
\vec r_i$
connecting the particle centers,  $\vec u  \cdot \vec r  >
0$, and the impact parameter $b = |\vec r  \times \vec u|/u $ is
smaller than the sum of the particle radii, $b \leq 
b_{\rm max}=d$.
If the particles repel each other, the
maximum impact parameter $b_{\rm max}$ becomes smaller than
$d$. By the conservation laws for angular momentum
and for energy one gets:
\begin{equation} \label{bmax}
b_{\rm max}^2 = d^2 \left( 1 - \frac{2 E_{\rm q}}{\mu \, u^2}
\right) \; ,
\end{equation}
where $\mu = m/2$ is the reduced mass.  $E_{\rm q}$ denotes the
energy barrier which must be overcome to let two particles collide
in the dilute limit. It is the
difference of the potential energies at contact and
when they are infinitely far apart.  Eq.~(\ref{bmax}) is
independent of the actual form of the potential, as long as it has
radial symmetry. (Note that energy is conserved as long as the
particles do not touch each other.)

Imagine a beam of particles, all having the same asymptotic velocity
$\vec u $ far away from particle $j$. All particles within an
asymptotic cylinder of radius $b_{\rm max}$ around the axis through
the center of $j$ with the direction of $\vec u $ will collide with
particle $j$. There will be $\pi \, b_{\rm max}^2 \, u \, n$ such
collisions per unit time, where $n=N/V$ is the number density.
Integrating over all relative velocities $\vec u$ gives the
collision frequency of a single particle in the granular gas in mean
field approximation:
\begin{equation}
  f = \pi n \int \limits_{u  \geq u_0} 
  d^3\!u \, u \, b_{\rm max}^2(u) \, p(u ) \; .
\end{equation}
$u_0 = \sqrt{2 E_{\rm q}/\mu}$ is the minimal relative velocity at infinity, for which a
collision can occur overcoming the repulsive interaction. We assume
that the particle velocity distribution is Gau{\ss}ian with variance $3
T$ (see (\ref{eq1})), so that the
relative velocity will have a Gau{\ss}ian distribution $p(u)$ as well, with
\begin{equation}
\langle u^2 \rangle= 6 \, T \; .
\end{equation}
Hence, the total number of binary collisions per unit time and per
unit volume is given by:
\begin{equation} \label{Ng}
\dot N_g = 1/2\, f\, n =
2\sqrt{\pi} \, n^2 \, d^2 \, \sqrt{T} \cdot
\exp\left(- \frac{E_{\rm q}}{mT} \right) \; .
\end{equation}
The factor $1/2$ avoids double counting of collisions. This
corresponds to textbook physics for chemical reaction rates as can be
found for example in Present\cite{Present}.

Now we calculate the dissipation rate: The energy loss due to a single
inelastic collision is:
\begin{equation}
\delta \!E(u, b) = \frac{\mu}{2} \, \left(1-e_{\rm n}^2\right) \,
{u_{\rm n}^*}^2 \; ,
\end{equation}
where $u_{\rm n}^{*}$ means the normal component of the relative
velocity $\vec u^{*}$ at the collision. It can be calculated easily 
from ${u_{\rm n}^{*}}^2={u^{*}}^2 - {u_{\rm t}^{*}}^2$: The tangential
component is determined by angular momentum conservation,
\begin{equation}
\mu u b = \mu u_{\rm t}^{*} d \; ,
\end{equation}
and energy conservation gives
\begin{equation}
{u^*}^2  =  u^2 
\left( 1 - \frac{2 E_{\rm q}}{\mu \, u^2} \right) =
u^2\left(\frac{b_{\rm max}}{d}\right)^2 \; .
\end{equation}
This yields
\begin{equation}
{u_{\rm n}^{*}}^2 = u^2
\frac{b_{\rm max}^2 -b^2}{d^2}  \; .
\end{equation}
The energy loss in one collision is therefore:
\begin{equation}
\delta \!E (u,b) = \frac{\mu}{2}  \, \left(1-e_{\rm n}^2\right) \, u^2 \,
\frac{b_{\rm max}^2 -b^2}{d^2}  \; .
\end{equation}
Assuming a homogeneous distribution of particles, we eliminate
the $b$-dependence by averaging over the area $\pi b_{\rm max}^2$:
\begin{eqnarray}
\delta \!E (u) & = & \frac{1}{\pi b_{\rm max}^2} 
\int \limits_{0}^{b_{\rm max}} db \, 2\pi b \, \delta \!E (u,b)
\\
 & = & \frac{\mu}{4} \, u^2 \, \left(1-e_{\rm n}^2\right)
\left( 1 - \frac{2 E_{\rm q}}{\mu \, u^2} \right) \; .
\label{eq11b}
\end{eqnarray}
The dissipated energy per unit time due to collisions with relative
velocity $u$ is then the number of such collisions per unit volume,
${1}/{2} \, n^2 \pi b_{\rm max}^2 \, u$, times the energy loss $\delta
\!E$, Eq.~(\ref{eq11b}).

Finally we get the dissipation rate per unit volume in the dilute
limit ($\nu \to 0$) by integration over the relative velocity
distribution:
\begin{eqnarray} 
\gamma & = & \frac{\pi}{2} n^2
\int \limits_{u \geq u_0} d^3u \,
b_{\rm max}^2 \, u \, \delta \!E (u) \, p(u)
\nonumber
\\
\label{gamma_tot}
 & = & 2\sqrt{\pi} \, n^2 d^2 m \, \left(1-e_{\rm n}^2\right) \,
T^{3/2} \cdot
\exp\left(- \frac{E_{\rm q}}{m T}\right) \; .
\end{eqnarray}
The dissipation rate of an {\em uncharged\/} granular system in the
dilute limit in the kinetic regime is given
by~\cite{Haff}:
\begin{equation} \label{gamma_0}
\gamma_0 = 2\sqrt{\pi} \, n^2 d^2 m \, \left(1-e_{\rm n}^2\right) \,
T^{3/2} \; .
\end{equation}
Thus the dissipation rate (\ref{gamma_tot}) in a dilute granular gas
with repulsive pair interactions and the one for the uncharged case
differ only by a Boltzmann factor,
$\gamma = \gamma_0 \cdot \exp\left(- {E_{\rm q}}/{m T}\right)$. 
This is the main result of the analytic treatment in this section.

If all grains carry the charge $q$, the potential barrier is $E_{\rm
  q} = q^2/d$. Our simulation results, Fig. \ref{work.fig.2}, show that
(\ref{gamma_tot}) is applicable in the dilute case at least for
$E_{\rm q}/mT < 10$. For smaller granular temperature, however, we
expect that corrections to (\ref{gamma_tot}) due to the long range of
the Coulomb potential become important.

\section{Dissipation rate for dense systems} \label{sec_DELTA}
In order to discuss the dissipation rate $\gamma$ in a non-dilute
system of charged granular matter, let us recall the analytic form of
$\gamma$ in an uncharged non-dilute system. The derivation is basically
done by using the Enskog expansion of the velocity distribution
function for dense gases~\cite{Lun84}. One finds that (\ref{gamma_0})
underestimates the collision rate (hence also the dissipation rate)
because it does not take the excluded volume into account. The 
corrected dissipation rate for a non dilute uncharged system is:

\begin{equation} \label{gamma_unch.dns}
  \gamma = \gamma_0 \cdot g_{\rm hs}(\nu) \; ,
\end{equation}
where $\gamma_0$ is given by Eq.~(\ref{gamma_0}) and $g_{\rm hs}(\nu)>1$ is
the equilibrium pair distribution function of the non-dissipative
hard-sphere fluid at contact. It only depends on the solid fraction
$\nu = \pi n d^3/ 6$:
\begin{equation} \label{g_hs}
g_{\rm hs}(\nu) = \frac{2-\nu}{2 (1-\nu)^3}
\end{equation}
 (Carnahan and
Starling~\cite{Carnahan}, Jenkins and Richman \cite{Jenkins}).

Our system consists of dissipative charged hard-spheres (CHS).  The
Boltzmann factor in Eq.~(\ref{gamma_tot}) is just the equilibrium pair
distribution function at contact in the dilute limit for a CHS-fluid,
$\lim_{\nu \to 0} g_{\rm chs}(\nu,q) = \exp \left(- E_{\rm q}/m T\right)$. So it
is plausible, that the dissipation rate for a dense system of
dissipative CHS should be
\begin{equation} \label{gamma_ch.dns}
  \gamma = \gamma_0 \cdot g_{\rm chs}(\nu, q) \; .
\end{equation}

Unfortunately the literature is lacking a satisfying analytic
expression for $g_{\rm chs}$. In 1972 Palmer and Weeks~\cite{Palmer72} did
a mean spherical model for the CHS and derived an analytic expression
for $g_{\rm chs}$, but this approximation is poor for low densities.
Many methods~\cite{bunch_of_chs}
give $g_{\rm chs}$ as a result of integral equations, that can be solved
numerically. We do not use those approximations, but
make the following ansatz for $g_{\rm chs}$:
\begin{equation} \label{gamma_ansatz}
  g_{\rm chs}(\nu,q) \approx g_{\rm hs}(\nu) \cdot \exp\left(-
  \frac{E_{\rm eff}(\nu)}{m T}\right) \; .
\end{equation}
As in the dilute case we assume that the long range Coulomb repulsion
modifies the pair distribution function of the uncharged hard sphere
gas by a Boltzmann factor. Note that the granular temperature enters
the pair distribution function only through this Boltzmann factor. The 
hard core repulsion is not connected with any energy scale, so that
the pair distribution function $g_{\rm hs}$ cannot depend on $T$.  
The effective energy barrier $E_{\rm eff}$ must approach $E_{\rm q}=q^2/d$ in the
dilute limit. Hence the ansatz (\ref{gamma_ansatz}) contains both the
uncharged and the dilute limit, (\ref{gamma_unch.dns}) respectively
(\ref{gamma_tot}). 

\begin{figure}[b]
\psfig{figure=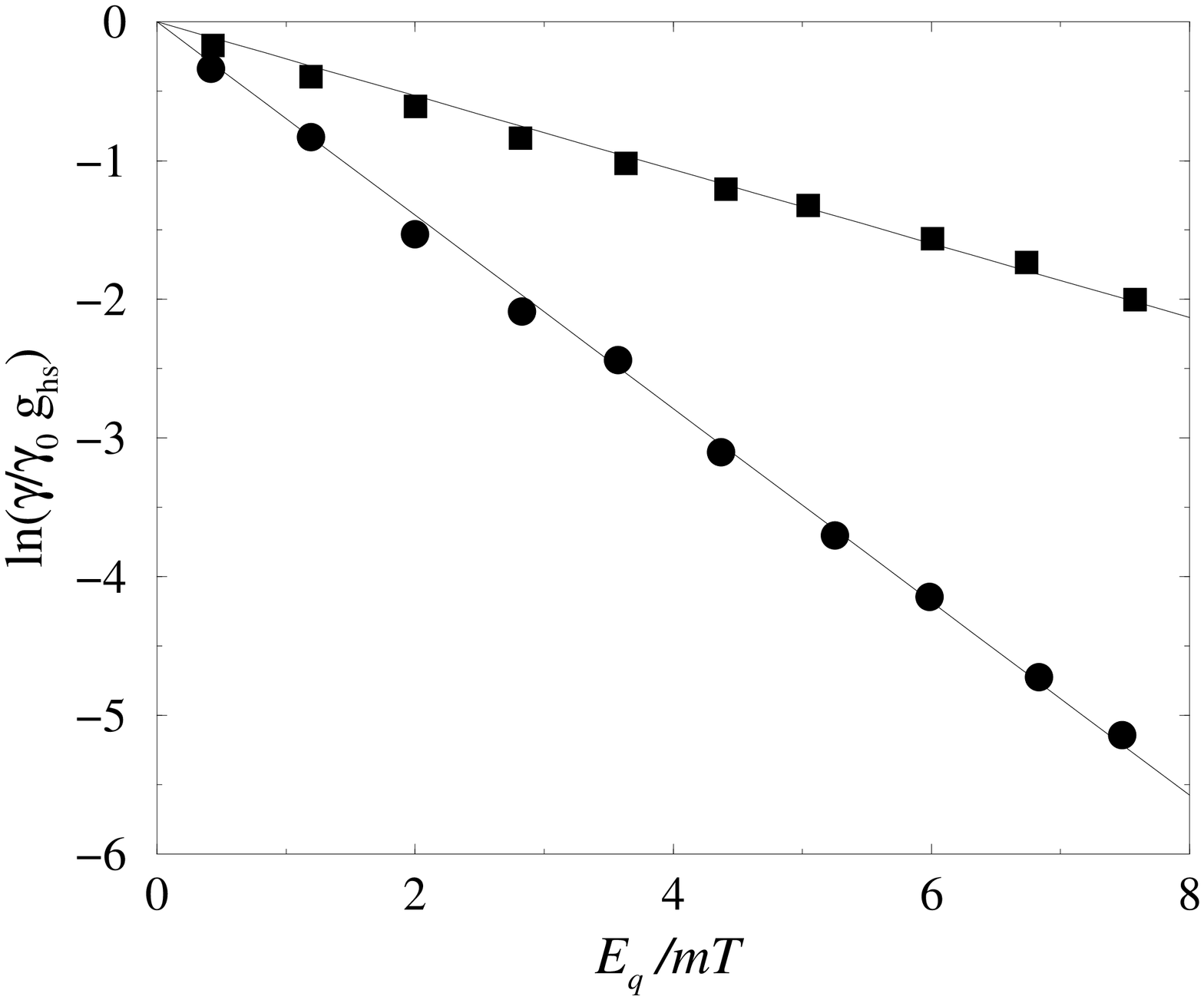,angle=270,width=8cm}
\caption{Arrhenius-plot of the dissipation rate $\gamma$ normalised by
  the one of the uncharged system,
  Eq.~(\ref{gamma_unch.dns}). Granular temperature is scaled by $E_{\rm q}/m$.
  Filled circles correspond to simulations of density $\nu = 3.375
  \cdot 10^{-3}$ and filled squares $\nu = 7 \cdot 10^{-2}$.  The
  linear fits yield: $E_{\rm eff}/E_{\rm q} = 0.70$ for the lower
  density and $E_{\rm eff}/E_{\rm q} = 0.27$ in the other case.}
\label{work.fig.2}
\end{figure}

In order to check the ansatz (\ref{gamma_ansatz})
we did computer simulations using the MD algorithm as described in the
appendix \cite{Scheffler_MD}. Test systems of varying solid fraction $\nu$ and particle
number ranging from $N=256$ to $N=1024$ were prepared at a starting
temperature $T_0$. As soon as the simulation starts, the granular
temperature drops because of the inelastic collisions. We measured the
dissipation rate $\gamma$ and the granular temperature during this
evolution. According to Eq.~(\ref{gamma_ansatz}) and
Eq.~(\ref{gamma_ch.dns}) the dissipation rate is $\gamma = \gamma_0 \,
g_{\rm hs}(\nu) \cdot \exp(E_{\rm eff}(\nu) /mT)$. An
Arrhenius plot ($\ln(\gamma/\gamma_0\, g_{\rm hs})$ versus $E_{\rm q}/mT$)
should give a straight line whose negative slope is the effective
energy barrier $E_{\rm eff}$. 
Fig.~\ref{work.fig.2} shows two examples of these simulations. The
Arrhenius plots are linear to a very good approximation. This confirms
the ansatz (\ref{gamma_ansatz}).
Systems with high densities show slight deviations from linearity.


The negative slopes $E_{\rm eff}/E_{\rm q}$ in Fig.~\ref{work.fig.2} are
smaller than $1$, which means, that the effective energy barrier is
smaller than in the dilute system. The explanation is that two
particles which are about to collide not only repel each other but
are also pushed together by being repelled from all the other charged
particles in the system. 

For dimensional reasons the effective energy barrier to be overcome,
when two particles collide, must be of the form
\begin{equation} \label{Phi}
E_{\rm eff}  = \frac{q^2}{d} - \frac{q^2}{\ell} f(d/\ell) \; ,
\end{equation}
where $\ell>d$ is the typical distance between the charged particles
and $f$ is a dimensionless function.
The first term is the Coulomb interaction $E_{\rm q}$ of the collision
partners at contact. The second term takes the interaction with all
other particles in the system into account. It is negative, because
the energy barrier for the collision is reduced in dense systems.

Obviously, for a dense packing, $\ell \rightarrow d$, the energy
barrier for a collision must vanish, i.e.
$ E_{\rm eff}|_{d=\ell} = 0 $.
Moreover, if one takes a dense packing and reduces the radii of all
particles infinitesimally, keeping their centers in place, all
particles should be force free for symmetry reasons. Therefore, the
energy barrier must vanish at least quadratically in $(\ell - d)$, i.e.
$\partial E_{\rm eff}/\partial d |_{d=\ell} =0$.
For the function $f$ this implies
\begin{equation}
  \label{f_assumptions}
    f(1) = 1 \qquad \text{and} \qquad
  \frac{{\rm d}f(x)}{{\rm d}x}\Bigg|_{x=1} = -1.
\end{equation}
If the particle diameter $d$ is much smaller than the typical distance
$\ell$ between the particles, the function $f(d/\ell)$ may be expanded
to linear order,
\begin{equation} \label{Taylor}
f(x) = c_0 + c_1 x + \ldots
\end{equation}
In linear approximation the coefficients are determined by
(\ref{f_assumptions}): $c_0 = 2$ and $c_1 = -1$.
This determines the energy barrier (\ref{Phi}).

In 1969 Salpeter and Van Horn~\cite{Salpeter69,Slattery80} pointed
out, that inside a 
strongly coupled OCP a short-range body centered cubic (BCC) ordering
will emerge. In the BCC lattice the nearest neighbour distance $\ell$
is related to the volume fraction $\nu$ by
\begin{equation}
\frac{d}{\ell} = \frac{2}{\sqrt{3}} \left(\frac{3}{\pi} \nu
\right)^{1/3}  \approx 1.14\,\nu^{1/3}.
\label{value}
\end{equation}
Assuming a BCC structure and using the linear approximation for $f(x)$
in (\ref{Phi}), the effective energy barrier is therefore given by
\begin{equation}
  \label{E_eff_bcc}
  E_{\rm eff} = E_{\rm q} \left(1-\frac{d}{\ell}\right)^2  
  = E_{\rm q} \left(1-2.27\,\nu^{1/3} + 1.29\,\nu^{2/3}\right)
\end{equation}

\begin{figure}[tb]
\psfig{figure=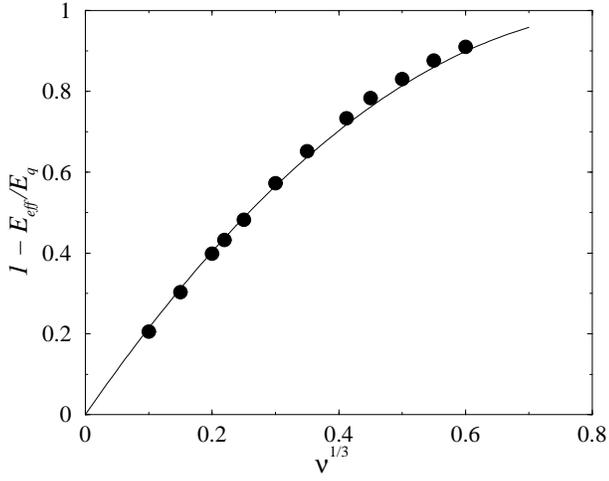,angle=270,width=8cm}
\caption{The dependence of the effective energy barrier on the solid
  fraction. Filled circles correspond to computer simulations, the
  solid line is Eq.~(\ref{E_eff_bcc}).}
\label{work.fig.3}
\end{figure}

To test Eq.~(\ref{E_eff_bcc}) we simulated systems with
densities ranging from $\nu = 0.001$ to $\nu = 0.216$ and determined
the ratio $E_{\rm eff} (\nu) / E_{\rm q}$ as in
Fig.~\ref{work.fig.2}. The results are plotted in
Fig.~\ref{work.fig.3}.  The agreement of the theoretical formula
(\ref{E_eff_bcc}) with the simulations is excellent. One can see, that
in the dilute limit the 
effective energy barrier extrapolates to $E_{\rm q}$. We cannot
simulate systems with very low density, because collisions are too unlikely. 

For the highest densities one cannot expect that the linear
approximation (\ref{Taylor}) remains valid. Also, the dense packing of
spheres is achieved with an FCC (face centered cubic) rather than a
BCC ordering. This may be responsible for the systematic slight 
deviation from the theoretical curve in Fig.~\ref{work.fig.3}.
A more refined analysis \cite{Scheffler_Corr} of the pair distribution function leads to 
$d/\ell \approx 1.20 \nu^{1/3}$ instead of (\ref{value}). This fits
the data for large $\nu$ in Fig.\ref{E_eff_bcc} slightly better.

The reduction of the Coulomb repulsion was also found in the OCP, when it
was applied to dense stars~\cite{Salpeter69}. There the analogue of the
second term in (\ref{Phi}) is called the ``screening potential''
(somewhat misleadingly, as there is no polarizable counter charge and
hence no screening). Monte Carlo simulations~\cite{Brush} of the OCP 
were interpreted in terms of a linear ``screening
potential''~\cite{DeWitt73}, which corresponds to (\ref{Taylor}), and
the analogue of the conditions (\ref{f_assumptions}) also occurs in the
plasma context~\cite{Itoh}, although based on a different physical
reasoning. Corrections to the linear approximation are the subject of
current research~\cite{Rosenfeld}. However, applying these more
sophisticated forms of the ``screening potential'' of the OCP model
to dense charged granular gases seems
arguable as for higher densities the influence of the hard
spheres become more and more important and so the analogy to the OCP
model, which uses point charges, does no longer hold.

We showed that (\ref{gamma_ch.dns}) and (\ref{gamma_ansatz}) also hold
in two dimensions with \cite{JR85,H}
\[
\gamma_0^{\rm 2D} = \sqrt{\pi} \, n^2 d m \, \left(1-e_{\rm n}^2\right)
T^{3/2}  , 
\] 
\begin{equation}
g_{\rm hs}^{\rm 2D}(\nu) = (1-\frac{7}{16}\nu)/(1-\nu)^2  \quad {\rm and}
\label{2dGamma}
\end{equation} 
\[
E_{\rm eff} = E_{\rm q} (1 - d/\ell)^2 \quad {\rm as\; in\; three\; dimensions}.
\]

%
\section{Time dependence of the granular temperature}\label{sec_COOL}

The results for the dissipation rate suggest that one must distinguish
two  asymptotic regimes for collisional cooling in a charged granular 
gas: As long as the granular temperature of the granular gas is so
high that the kinetic energy of the grains is much larger than the
Coulomb barrier $E_{\rm
  eff}$, the charges can be neglected. The collisional cooling then
proceeds initially like in the uncharged case, i.e. the granular temperature
decreases as $t^{-2}$. However, as the kinetic energy approaches $E_{\rm
  eff}$ the charges become more and more important. Electrostatic
repulsion suppresses the collisions, so that
the collisional cooling slows down dramatically. This will be
calculated in this section.

If one assumes that the dissipation of energy essentially changes only
the kinetic energy, it follows that:
\begin{equation} \label{col.1}
  \frac{d}{dt} E_{\rm kin}/V = - \gamma\left( T(t) \right)
\end{equation}
With $E_{\rm kin}/N = 1.5 m T$ this gives a
differential equation for $dT/dt$:
\begin{equation}
  \frac{dT}{dt} = - \frac{2}{3} m^{-1} n^{-1} \gamma\left(T(t)\right)
\end{equation}
This gives with
(\ref{gamma_0}), (\ref{gamma_ch.dns}) and (\ref{gamma_ansatz}):
\begin{equation}\label{dT/dt}
  \frac{dT}{dt} = - \frac{4}{3} \sqrt{\pi} n d^2 (1 - e_{\rm n}^2)
  T^{3/2} \cdot g_{\rm hs} \cdot \exp \left( -
  \frac{E_{\rm eff}}{m T} \right)
\end{equation}

As the total number of collisions is an increasing function in
time, we can choose the number of collisions of particles, $c :=
\mathrm{collisions} / N$, as a measure of time. This substitution is
known from uncharged granular cooling~\cite{Goldhirsch2}. We get
\begin{equation}
  \frac{dT}{dc} = \frac{dT}{dt}/\frac{dc}{dt} = 
-\frac{2}{3} \left(1-e_{\rm n}^2\right) T(c)
\end{equation}
where the derivative
\begin{equation}
  \frac{dc}{dt} = 2 \sqrt{\pi} n d^2 \sqrt{T} \cdot g_{\rm hs} \cdot
  \exp \left( - \frac{E_{\rm eff}}{mT} \right)
\end{equation}
is derived from (\ref{Ng}), (\ref{g_hs}) and (\ref{gamma_ansatz}).
This gives the solution
\begin{equation} \label{col.2}
  T(c) = T_0 \cdot \exp \left(- \frac{2}{3} \left(1 - e_{\rm n}^2 \right)
  c \right)
\end{equation}
where $T_0$ is the temperature, where the counting of collisions starts.

\begin{figure}[tb]
  \begin{center}
    \psfig{figure=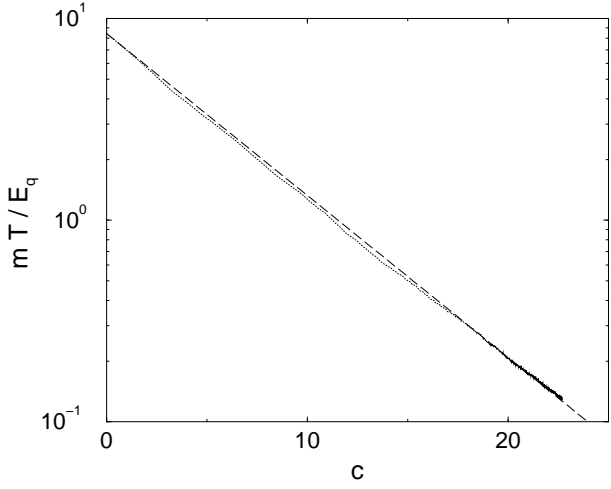,angle=270,width=8cm}
\caption{Cooling of a sytem with density $\nu = 3.375\cdot 10^{-3}$
  and $e_{\rm n} = 0.85$. The temperature is normalised by $E_{\rm
    q}/m$ and $c$ is the number of collisions per particle. The dashed
    line is given by (\ref{col.2}).}
\label{coll.fig}
\end{center}
\end{figure}

Fig.~\ref{coll.fig} shows the free cooling of a test system. The dashed
line corresponds to the approximative solution of the theory given
above, and the solid curve is the result of a computer simulation. The
agreement between (\ref{col.2}) and the simulation is very good. This
means that the change of the overall potential energy can be neglected
compared to the change of the kinetic energy for high temperatures.

An analytical solution of (\ref{dT/dt}) is obtained by substituting
\begin{equation}
u=\sqrt{\frac{E_{\rm eff}}{m T}} \quad {\rm and}\quad
\tau = \left( \frac{2}{3} \sqrt{\pi} n d^2 (1-e_{\rm n}^2) g_{\rm hs}
\sqrt{\frac{E_{\rm eff}}{m}} \right) t.
\end{equation}
This gives
\begin{equation}
\frac{du}{d\tau} = \exp(-u^2)
\end{equation}
with the initial condition $u(0)= \sqrt{E_{\rm eff}/m T_0} $.
With the integral $I(u) = \int_0^u \exp(v^2) dv$, which is related to
the probability function $\Phi(x) = (2/\sqrt{\pi}) \int_0^x 
\exp(-\xi^2) d\xi$ by 
\begin{equation}
I(u) = \frac{\sqrt{\pi}}{2i} \Phi(iu),
\end{equation}
the time $t(T,T_0)$ during which the granular temperature drops from
$T_0$ to $T$ is given by
\begin{equation}
t(T,T_0) = \left( I\left(\sqrt{\frac{E_{\rm eff}}{mT}}\right)
- I\left(\sqrt{\frac{E_{\rm eff}}{mT_0}}\right)\right)\left/
C \sqrt{\frac{E_{\rm eff}}{m}}\right.
\label{t(T)}
\end{equation} 
with the constant
\begin{equation}
C = \frac{2}{3} \sqrt{\pi} n d^2 (1-e_{\rm n}^2) g_{\rm hs}.
\end{equation}

For granular temperatures which are large compared to the 
Coulomb barrier, $T_0 > T \gg E_{\rm eff}/m$, the right
hand side of (\ref{t(T)}) may be approximated as $ I'(0)
(T^{-1/2} - T_0^{-1/2})/C $. With $I'(0) = 1$ this reduces
to the well known equation for the free cooling of an uncharged,
homogeneous granular gas. For large initial granular temperature,
$T_0 \rightarrow \infty$, one obtains a power law
\begin{equation}
T \approx (C t)^{-2}.
\end{equation}
This power law is only valid for $t \ll t_{\rm c} = (E_{\rm
    eff}C^2/m)^{-1/2}$. At $t_{\rm c}$, the granular temperature $T$
    drops below $\sqrt{E_{\rm 
    eff}/m}$. Then the repulsion between the charges on the particles
becomes important. Using $I(u) \approx \exp(u^2)$ for
large $u$ one obtains for $t\gg t_{\rm c}$
\begin{equation}
T \approx \frac{E_{\rm eff}}{m}\frac{1}{\ln(t/t_{\rm c})}.
\end{equation}
%
%
\section{A few applications to heterogeneous systems}\label{sec_APPL}

\subsection{Influence of walls}
\label{Walls}

Due to the long range of unscreened Coulomb interactions, monopolarly
charged granular systems are always inhomogeneous in reality, because the
influence of container walls is not restricted to their vicinity. Therefore it
is an important question, whether results obtained for homogeneous systems
with periodic boundary conditions can be applied locally.

We checked this for a two dimensional charged granular gas confined in
$x$-direction by walls, which are assumed to be uniformly charged and 
unpolarizable for simplicity so that their electrical field inside the box is
zero. In $y$-direction periodic boundary conditions were imposed so that the
system is translationally invariant in this direction. 

\begin{figure}[tb]
\psfig{figure=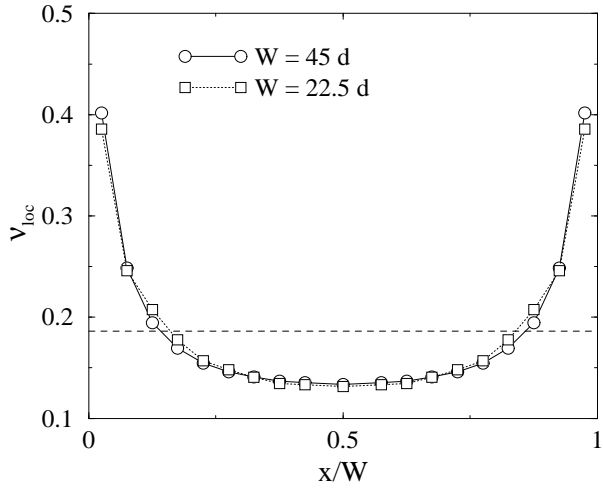,angle=0,width=8cm}
\caption{Solid fraction of a charged granular gas in a two dimensional
box with confining walls perpendicular to the $x$-direction and
periodic boundary conditions in the $y$-direction.
The dimensions of the box are $W\times L$ with $L= 60 d$ and two different
values of $W$. Because of Coulomb repulsion the particle concentration
increases towards the wall.  Because of the absence of screening the
solid fraction profile depends approximately only on the scaled
variable $x/W$,  with small deviations due to the excluded
volume interaction. }
\label{solidfraction}
\end{figure}

The box was divided into
equal layers parallel to the walls, and the properties were averaged over these
layers. In this way the solid fraction shown in Fig.\ref{solidfraction} was
obtained. As expected, the particle concentration increases towards the walls,
because there is no Coulomb repulsion from particles outside the container.
As screening is absent inside the box the solid fraction profile for
fixed global solid fraction, $\nu = 0.186$ (averaged over the entire 
container), depends approximately only on the scaled variable $x/W$, where
$W$ is the distance between the walls, with small deviations due to the 
excluded volume interaction. 

Similarly we recorded the local granular temperature $T_{\rm loc}
(x/W)$ (not shown) and the local dissipation rate, $\gamma_{\rm loc}
(x/W)$. All these profiles depend on the global granular temperature, $T(t)$,
which decreases due to collisional cooling. However, in order to get
good statistics, we set the coefficient of restitution 
equal one in this simulation so that $E_{\rm q}/mT = 1.7$ remained
constant. Hence the local profiles of the solid
fraction and the granular temperature could be averaged over time. The
local dissipation rate was calculated by taking for each collision the
energy into account, which would have been dissipated if the
coefficient of restitution had been zero. 

\begin{figure}[tb]
\psfig{figure=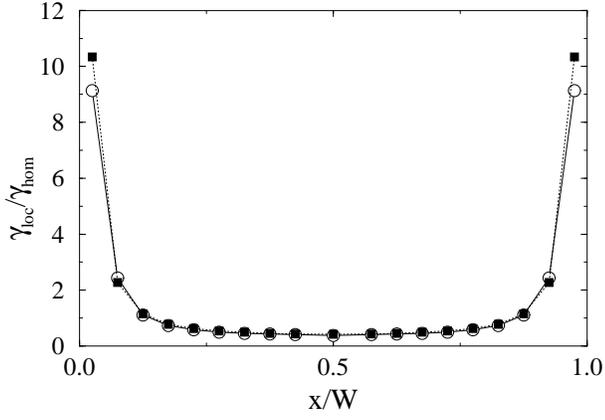,angle=0,width=8cm}
\caption{Same system as Fig.\ref{solidfraction} with $W=45 d$. Open circles
are the local dissipation rates, in units of
the dissipation rate in a homogeneous system with solid fraction $\nu=0.186$
at the average granular temperature in the box, $E_{\rm q}/mT = 1.7$. 
The black squares are the dissipation rates calculated from
(\ref{2dGamma}) with the local solid fraction (see Fig.\ref{solidfraction}) and
the local granular temperature.  }
\label{gamma.profile}
\end{figure}

Fig.\ref{gamma.profile} shows the local dissipation rate in units of 
$\gamma_{\rm hom}$, the dissipation rate  
for a homogeneous system with the same global solid fraction
and the same global temperature. It agrees very well with the
theoretical prediction obtained from (\ref{gamma_ch.dns}),
(\ref{gamma_ansatz}) and (\ref{2dGamma}), if one inserts 
the local solid fraction, $\nu_{\rm loc}$ (see Fig.\ref{solidfraction}), and
the local granular temperature, $T_{\rm loc}$. Apart from the
immediate neighborhood of the wall the system is locally
homogeneous enough that the results from the previous sections may be applied.

\subsection{Pipe flow}

\begin{figure}[tb]
\psfig{figure=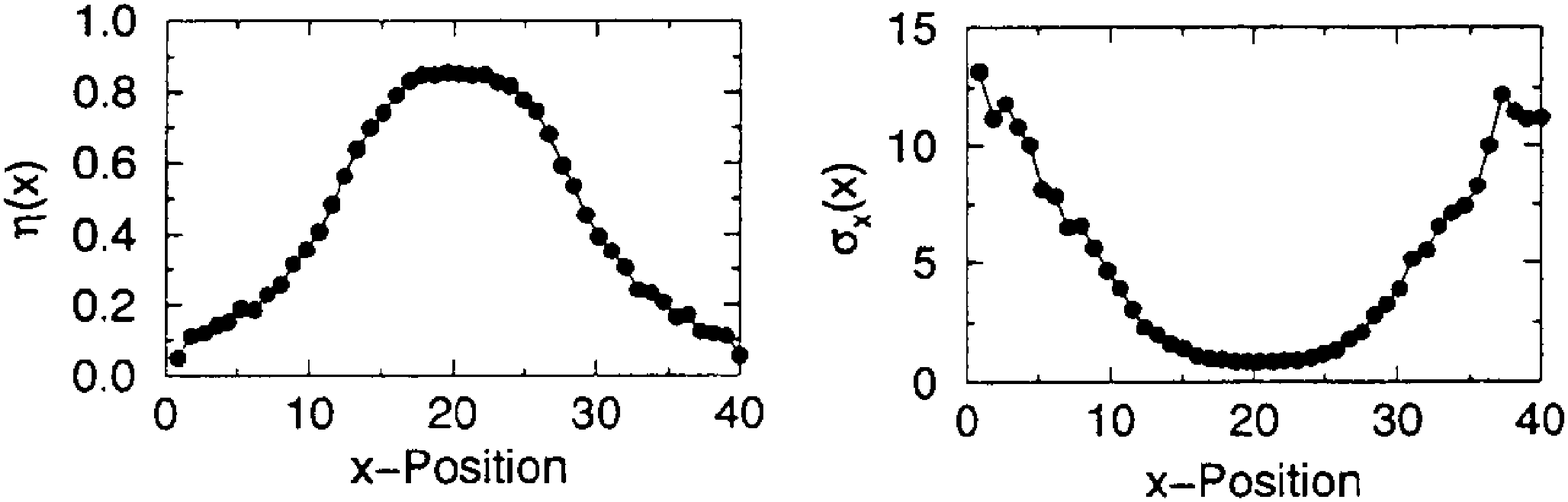,angle=0,width=8cm}
\caption{Local solid fraction $\eta$ (left) and root mean square fluctuation of the
  horizontal velocity component (right) for granular flow through a
  two dimensional vertical pipe. The quantities are averaged over
  layers parallel to the wall in order to show their
  $x$-dependence. The diameter of the pipe is 40 $d$. From \cite{SchaeferDiss}. }  
\label{7.5}
\end{figure}

The flow of granular matter through a vertical pipe depends strongly
on whether or not the particles are charged. First we recall the
results for the uncharged case \cite{Wolf99,SchaeferDiss}. Steady state flow
is reached, when gravitational acceleration and friction at the walls
balance each other. In this case, ``friction'' is due to collisions of
grains with the wall, by means of which momentum is transfered and the
particles are randomly scattered back. Hence the walls are permanent
sources of granular temperature, which are balanced by the collisional
cooling within the pipe. Fig. \ref{7.5} shows typical steady state
profiles of local volume fraction and local granular temperature across 
a pipe: The granular gas is dilute close to the walls, where the
granular temperature is high, and gets compressed towards the middle
of the pipe, where the granular temperature is low. For this
inhomogeneity the pipe must not be too narrow.

Comparing this with Fig.\ref{solidfraction}, it is obvious, that 
electrostatic repulsion counteracts the dilation near the wall
and the compression in the interior of the pipe: If all particles
carry charges of the same sign, the solid fraction profile will be
flatter than in the uncharged case. The increased concentration near
the walls will lead to more ``friction'' at the wall. Hence charged 
matter flows more slowly through a vertical pipe and may even get stuck 
for strong charging \cite{Aider99}.

\subsection{Differently charged particles}

In a real system the particles will carry different charges, even if
they are equal in all other respects. The charging process may
determine the sign of the charges, but the amount will in general
fluctuate. Therefore it is important to discuss how the dissipation
rate depends on the width of a charge distribution, if the total
charge of the system is fixed.

First we consider that half of the particles carry a charge 10 times
larger than the other half: $q_{\rm B} = 10 q_{\rm A}$. 
The energy barriers for collisions between two A-particles, an A- and
a B-particle, or  two B-particles are now different:
\begin{equation}
E_{\rm AA} = \frac{q_{\rm A}^2}{d} < E_{\rm AB} = \frac{q_{\rm A}
  q_{\rm B}}{d} < E_{\rm BB} = \frac{q_{\rm B}^2}{d} \; .
\end{equation}

As A-particles repell each other much less, one expects them to
collide more frequently than B-particles. Remarkably the granular
temperature of the A-particles and the B-particles remained the same
in our simulations: A-particles that have lost part of their kinetic
energy in a collision are stirred up by the strongly charged particles.
The redistribution of kinetic and potential energy
happens so quickly that the weakly and the strongly charged particles
do not decouple thermally. This is in contrast to uncharged systems
\cite{Barrat02}.

\begin{figure}[tb]
\psfig{figure=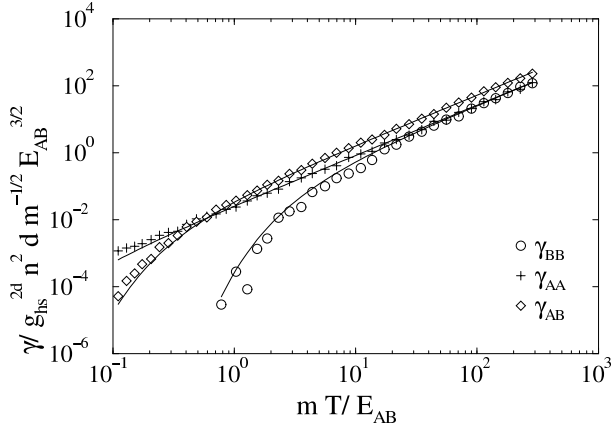,angle=0,width=8cm}
\caption{Dissipation rates $\gamma_{\alpha\beta}$ as a
  function of granular temperature in
  convenient units. There are equal numbers of particles A and B in 
  the system, which differ only by their charges, $q_{\rm A} = 0.1
  q_{\rm B}$. At high granular temperatures the dissipation rates due
to A-A, B-B, or A-B collisions, respectively, obey a $T^{3/2}$ power
  law, where $\gamma_{\rm AB}$ is larger by a factor of 2 compared to
the other two curves, because $p_{\rm AA} = p_{\rm BB} = p_{\rm AB}/2$.
For decreasing $T$ first the collisions between the strongly charged
  B-particles are suppressed. For $E_{\rm AA}/E_{\rm AB} \ll mT/E_{\rm
  AB} \ll 1$ only 
  collisions between A-particles still contribute to the dissipation
  rate. For still smaller granular temperature (not shown) also this
contribution will be suppressed. The curves are given by the
  theoretical formula (\ref{gamma_AB}).}
\label{fig6.5}
\end{figure}

Based on the results for uniformly charged particles we postulate 
that the dissipation rate due to collisions between particles 
$\alpha$ and $\beta$ (A or B) is approximately given by
\begin{equation}
\gamma_{\alpha \beta} = \gamma_0 \cdot g_{\rm hs}\cdot p_{\alpha \beta}
\cdot \exp{\left( -\frac{E_{\alpha\beta}}{mT} \left( 1 -
      \frac{d}{\ell}\right)^2 \right)} \; , 
\label{gamma_AB}
\end{equation}
where $p_{\rm AA}= c_{\rm A}^2$, $p_{\rm AB}= 2 c_{\rm A}c_{\rm B}$ and $p_{\rm
  BB}= c_{\rm B}^2$ with the concentration $c_{\rm A} = 1 - c_{\rm B}$
  of A-particles (in our simulation $c_{\rm A}=c_{\rm B}=1/2$). Fig.\ref{fig6.5} shows
  a comparison of (\ref{gamma_AB}) with simulation results for a
  two dimensional system with solid fraction $\nu = 0.09$ and
  coefficient of restitution $e_{\rm n}= 0.97$. The agreement for high
  granular temperatures is excellent. For low granular temperatures
  there are slight systematic deviations. These deviations are
  probably due to the fact that the Ansatz for the effective Coulomb barrier 
$E_{\alpha\beta} \left( 1 - \frac{d}{\ell}\right)^2$ in
  (\ref{gamma_AB}) is too naiv.

Now we apply (\ref{gamma_AB}) to compare the dissipation rates 
$\gamma = \gamma_{\rm AA} + \gamma_{\rm AB} + \gamma_{\rm BB}$ for
systems with general bimodal charge distributions, where all charges
have equal sign. Two independent parameters characterize the charge distribution
for given total charge $N\bar q$ in the system with $N$
particles. These are the concentration $c_{\rm A}$ of the weakly
charged particles and their charge, $q_{\rm A}$. The concentration and
charge of the strongly charged particles are given by
\begin{equation}\label{c_B,q_B}
c_{\rm B} = 1 - c_{\rm A}, \quad q_{\rm B} = \frac{\bar q - c_{\rm
    A}q_{\rm A}}{c_{\rm B}}.
\end{equation}
As the physics does not depend on the sign of the charges, as long as it is 
the same for all particles, we may assume in the following that 
$0 \leq q_{\rm A} \leq \bar q$. The concentration parameter is
restricted to the interval $0 \leq c_{\rm A} < 1$.

We now show, that 
\begin{equation}
\frac{\partial \gamma (c_{\rm A},q_{\rm A})}{\partial c_{\rm A}} \geq 0
\label{variation1}
\end{equation}
and
\begin{equation}
\frac{\partial \gamma (c_{\rm A},q_{\rm A})}{\partial q_{\rm A}} \leq
0 \, .
\label{variation2}
\end{equation}
Since uniformly charged systems correspond to the lower boundary of
the concentration interval, $c_{\rm A}=0$, or the upper boundary of
the charge interval, $q_{\rm A}= \bar q$, repectively,
(\ref{variation1}) and (\ref{variation2}) imply that for a given
average charge $\bar q$ the lowest collision rate (or dissipation
rate) is reached, if all particles carry the same charge, $bar q$.

In order to prove the inequalities (\ref{variation1}) and
(\ref{variation2}) it is convenient to introduce the normalized quantities 
\begin{equation}
\tilde\gamma_{\alpha\beta}=\gamma_{\alpha\beta}/\gamma_0 g_{\rm hs} \quad
, \qquad
\tilde q_{\alpha}=(q_{\alpha}/\sqrt{dmT})(1-d/\ell)
\end{equation}
so that
$\tilde\gamma_{\alpha\beta}=p_{\alpha\beta}\exp(-\tilde q_{\alpha}\tilde q_{\beta})$.
Using (\ref{c_B,q_B}) one obtains
\begin{equation}
\frac{\partial\tilde\gamma}{\partial c_{\rm A}}
= 2 c_{\rm A} e^{-\tilde q_{\rm A}^2}
+ 2 \left( c_{\rm B} - c_{\rm A} 
- c_{\rm A}c_{\rm B}\tilde q_{\rm A}\frac{\partial \tilde q_{\rm B}}{\partial c_{\rm A}} \right) 
  e^{-\tilde q_{\rm A}\tilde q_{\rm B}}     
- 2 c_{\rm B} \left(1+ c_{\rm B}\tilde q_{\rm B}\frac{\partial \tilde q_{\rm B}}{\partial c_{\rm A}} \right)
e^{- \tilde q_{\rm B}^2}.
\end{equation}
Inserting
$\partial \tilde q_{\rm B}/\partial c_{\rm A} = (\tilde q_{\rm B} - \tilde q_{\rm A})/c_{\rm B}$
%
%
this can be written in the form
\[
\frac{e^{\tilde q_{\rm A}^2}}{2 c_{\rm A}} \frac{\partial \tilde\gamma}{\partial c_{\rm A}}
= \left[ 1-\left(1+\tilde q_{\rm A}(\tilde q_{\rm B}-\tilde q_{\rm
  A})\right) e^{-\tilde q_{\rm A}(\tilde q_{\rm B}-\tilde q_{\rm A})}
\right]
\]
\begin{equation}
+ \frac{c_{\rm B}}{c_{\rm A}} e^{-\tilde q_{\rm A}(\tilde q_{\rm B}-\tilde q_{\rm A})}
\left[ 1-\left(1+\tilde q_{\rm B}(\tilde q_{\rm B}-\tilde q_{\rm A})\right) e^{-\tilde q_{\rm B}(\tilde q_{\rm B}-\tilde q_{\rm A})} \right] .
\label{dGamma_dc}
\end{equation}
As $1+\tilde q_{\alpha}(\tilde q_{\rm B}-\tilde q_{\rm A}) 
\leq \exp(\tilde q_{\alpha}(\tilde q_{\rm B}-\tilde q_{\rm A}))$ ,
%
%
both square brackets and hence the whole expression (\ref{dGamma_dc}) are non-negative, as asserted in 
(\ref{variation1}). 

Similarly one calculates
\begin{equation}
\frac{\partial \tilde\gamma}{\partial \tilde q_{\rm A}} 
= -2 c_{\rm A}^2 \tilde q_{\rm A} e^{-\tilde q_{\rm A}^2} 
  -2 c_{\rm A}c_{\rm B} (\tilde q_{\rm B}
  + \tilde q_{\rm A}\frac{\partial \tilde q_{\rm B}}{\partial \tilde q_{\rm A}} ) 
  e^{-\tilde q_{\rm A}\tilde q_{\rm B}} 
  -2 c_{\rm B}^2 \tilde q_{\rm B}\frac{\partial \tilde q_{\rm B}}{\partial \tilde q_{\rm A}} 
  e^{-\tilde q_{\rm B}^2}.
\end{equation}
Using $\partial \tilde q_{\rm B}/\partial \tilde q_{\rm A} = - c_{\rm A}/c_{\rm B}$
this becomes
\begin{equation}
- \frac{1}{2 c_{\rm A}} \frac{\partial \tilde\gamma}{\partial \tilde q_{\rm A}} 
=  c_{\rm A} \tilde q_{\rm A} 
   \left(e^{-\tilde q_{\rm A}^2} - e^{-\tilde q_{\rm A}\tilde q_{\rm B}}\right)
 + c_{\rm B} \tilde q_{\rm B} 
   \left(e^{-\tilde q_{\rm A}\tilde q_{\rm B}} - e^{-\tilde q_{\rm B}^2}\right)
\label{dGamma_dq}
\end{equation}
As both terms on the right hand side are obviously positive this
proves (\ref{variation2}).

It is not difficult to prove along the same lines that 
\begin{equation}
\gamma_{\rm AB} \leq \gamma_{\rm AB_1} + \gamma_{\rm AB_2},
\end{equation}
if one splits up the B-fraction according to
$c_{\rm B}=c_{\rm B_1}+c_{\rm B_2}$ and $c_{\rm B}q_{\rm B}=c_{\rm
  B_1}q_{\rm B_1} + c_{\rm B_2}q_{\rm B_2}$. Moreover, the above
result, that a uniform charge leads to a smaller
collision rate than any bimodal distribution, implies also
\begin{equation}
\gamma_{\rm BB} \leq \gamma_{\rm B_1 B_1} + \gamma_{\rm B_1 B_2} +
\gamma_{\rm B_2 B_2} \, .
\end{equation}
Likewise, the A-fraction can be split up, and the whole procedure can
be applied iteratively, so that we reach the conclusion that the
collision rate for any charge distribution is higher than for the
case, where all particles carry the average charge.

It should be noted, however, that a broader charge distribution with
fixed average charge $\bar q$ does not necessarily lead to a larger
collision rate. In order to show this, we return to the bimodal
distribution as an example. The variance of the charge distribution is
\begin{equation}
\sigma^2 (c_{\rm A},q_{\rm A}) = c_{\rm A}(q_{\rm A} - \bar q)^2 +
                               c_{\rm B}(q_{\rm B} - \bar q)^2  
                               = \frac{c_{\rm A}}{1-c_{\rm A}}(q_{\rm
                               A} - \bar q)^2  \, .
\end{equation}
The partial derivatives of $\gamma$, (\ref{variation1}) and (\ref{variation2}),
have the same signs as the corresponding partial
derivatives of $\sigma^2$:
\begin{equation} 
\frac{\partial\sigma^2}{\partial c_{\rm A}} = \left(\frac{\bar q -
    q_{\rm A}}{1-c_{\rm A}} \right)^2 \geq 0 ,\qquad
\frac{\partial\sigma^2}{\partial q_{\rm A}} = -2c_{\rm A} \left(\frac{\bar q -
    q_{\rm A}}{1-c_{\rm A}} \right) \leq 0 \; . 
\end{equation}
However, as
$\nabla \gamma = (\partial \gamma/\partial c_{\rm
  A}, \partial \gamma/\partial q_{\rm A})$ and $\nabla \sigma^2 =
(\partial \sigma^2/\partial c_{\rm A}, \partial \sigma^2/\partial
q_{\rm A})$ are not parallel,  one can always find parameter changes
$d{\bf p} = (d c_{\rm A}, d q_{\rm A})$ such that $d \gamma = \nabla
\gamma \cdot d{\bf p}$ and $d \sigma^2 = \nabla \sigma^2 \cdot d{\bf
  p}$ have opposite signs. 

%

\section{Discussion}\label{sec_DISC}

We derived the dissipation rate of a charged granular gas, where all
charges have the same sign. Compared to
the uncharged case the dissipation rate is exponentially 
suppressed by a Boltzmann factor depending on the ratio between the
Coulomb barrier and the granular temperature.

In the derivation we assumed a Gau{\ss}ian velocity distribution,
although it is known that in the 
uncharged case  deviations from a Gau{\ss}ian behaviour emerge due to the
inelastic collisions~\cite{Esipov}. These deviations, however, were
shown to have little effect on the dissipation rate~\cite{Noije98}.
As the system becomes less dissipative in our case, it is reasonable
to expect that the effect of deviations from a Gau{\ss}ian velocity
distribution will be even weaker. 

In a dense system particle correlations enter the collision statistics
and hence the dissipation rate in two ways: First there is the well
known Enskog correction as in the uncharged case. It describes that 
the excluded volume of the other particles enhances the probability
that two particles are in contact. Second the Coulomb barrier which
colliding particles must overcome is reduced and will vanish in the
limit of a dense packing.

These results were obtained for periodic boundary conditions, so that
the system remained homogeneous. We showed, however, that it may be
applied locally in the case, where walls induce inhomogeneous solid
fraction and granular temperature. This inhomogeneity is opposite
to the one induced by flow of uncharged grains through a vertical
pipe. Therefore we could conclude that the Coulomb repulsion reduces
the flow velocity.

Coulomb repulsion slows collisional cooling
down from the $t^{-2}$-decay to a behaviour like
$1/\ln{t}$, when $T(t)$ drops below $T_{\rm c} = E_{\rm eff}/m$,
which means that the average kinetic energy of the grains does not
suffice to overcome the Coulomb repulsion. 

Finally we considered a bimodal charge distribution (both charges of
equal sign), which leads to more frequent collisions  among the weakly
charged particles than among the strongly charged ones. Nevertheless
the two kinds of particles kept the same granular temperature. We
pointed out, how the dissipation rate for such a system can be
evaluated. If the charge gets redistributed among the particles
($q_{\rm A} \rightarrow q_{\rm A} + \delta q$,
$q_{\rm B} \rightarrow q_{\rm B} - (c_{\rm A}/c_{\rm B}) \delta q$)
for fixed concentrations, then the dissipation rate increases with
the variance of the charge distribution. In particular the dissipation
rate is minimal, if all particles have the same charge. More general
we concluded that any deviation from uniform charging in a system with
given total charge increases the collision rate. 

An important question, which remained open in this paper, concerns the
influence of charge transfer processes on the collision statistics. In
the present investigation all particles retained their charges, even
if there was a charge difference. Therefore our theory need to be
modified for metallic particles.

\section*{Acknowledgements}

We thank Lothar Brendel, Haye Hinrichsen, Zeno Farkas, Alexander
C. Schindler and Hendrik Meyer for useful comments.  We gratefully
acknowledge support by the Deutsche 
Forschungsgemeinschaft through grants No.~{Wo~577/1-2} and
Hi 744/2-1.

\appendix

\section{Computer Simulation Method} \label{sec_COMP}

Distinct element (or molecular dynamics (MD)) simulations~\cite{AT}
are usually done with {\em time step driven} or {\em event driven}
algorithms~\cite{Wolf}. None of them is well suited for investigating
a charged granular gas. Therefore we developed a new simulation
scheme, which combines the virtues of both and will be described in
this section.

We use a ``brute force'' MD algorithm, which is simple and sufficient
for our problem. More sophisticated ways  of dealing with the long
range interactions, such as the
multipolar expansion~\cite{Multipol},
the particle-particle-particle-mesh~\cite{Eastwood} or the hypersystolic
algorithms~\cite{Lippert} should be used, if larger systems need to be
studied.

The event driven method for simulating the motion of all particles in
the granular gas can be applied, whenever the particle trajectories
between collisions can be calculated analytically, so that the time
interval between one collision event and the next can be skipped
in the simulation. Obviously this is impossible in a system with long
range Coulomb interactions. However, the idea to avoid the detailed
resolution of a collision event in time is still applicable. So the
velocities of the collision partners are simply changed
instantaneously to the new values predicted by momentum and angular
momentum conservation and an energy loss determined by the restitution
coefficient. We shall keep this feature of event driven simulations.

In the time step driven simulation method the equations of motion of
all particles in the granular gas are discretized using a fixed time
step, which is small compared to the duration of a collision. Hence
each collision, which is modeled as an overlap between particles, is
temporally resolved in detail. This has the advantage, that the
formation of long lasting contacts between particles can in principle
be simultated realistically. If the particles carry equal charges,
however, this will not happen, so that the collisions may be
approximated as being instantaneous like in event driven simulations.
Apart from being more efficient, this automatically avoids the so
called brake-failure artifact~\cite{Schaefer}, which hampers time-step
driven molecular dynamics simulations with rapid relative motion.  On
the other hand, we need a time discretization of the particle
trajectories between collisions, in order to take the changing
electrostatic interactions properly into account.


Because of the long-range nature of the Coulomb potential, we have to
include the interactions with the periodic images of the particles in
the basic cell. One way to do this is by Ewald summation. Details
of this method can be found in~\cite{AT}. 
Another method is the {\em minimum image\/}
convention: Only the nearest periodic image is taken into account for
the calculation of the interactions. The minimum image method has the
advantage, that it is much faster than the Ewald summation.
We checked the validity of the minimum image method by comparison with
the Ewald summation and found, that as long as $E_{\rm q}/mT < 10$ both
methods yield indistinguishable results. This upper limit for the coupling has
been found before in Monte-Carlo simulations of the OCP~\cite{Brush}.
As our systems all satisfy this condition, we used the minimum image
convention.



\newpage

\end{document}